\documentclass{article}

\pdfoutput=1
\usepackage{arxiv}
\usepackage[super]{nth}
\usepackage{subcaption}
\usepackage{textcomp}
\usepackage[utf8]{inputenc} % allow utf-8 input
\usepackage[T1]{fontenc}    % use 8-bit T1 fonts
\usepackage{hyperref}       % hyperlinks
\usepackage{url}            % simple URL typesetting
\usepackage{booktabs}       % professional-quality tables
\usepackage{amsfonts}       % blackboard math symbols
\usepackage{nicefrac}       % compact symbols for 1/2, etc.
\usepackage{microtype}      % microtypography
\usepackage{graphicx}
\usepackage{graphics}
\usepackage{amsmath}

\title{Learning Real Estate Automated Valuation Models from Heterogeneous Data Sources}

\author{
 Francesco Bergadano \\
  Department of Computer Science \\ University of Turin\\
  Turin, Italy \\
  \texttt{fpb@di.unito.it} \\
  \And
  Roberto Bertilone \\
  Department of Computer Science \\ University of Turin\\
  Turin, Italy \\
  \texttt{r.bertilone@gmail.com} \\
  \And
  Daniela Paolotti \\
  ISI Foundation \\ 
  Turin, Italy \\
  \texttt{daniela.paolotti@isi.it} \\
 \And
 Giancarlo Ruffo \\
  Department of Computer Science \\ University of Turin\\
  Turin, Italy \\
  \texttt{ruffo@di.unito.it} \\
}

\begin{document}
\maketitle

% The abstract is a short summary of the work to be presented in the article.
\begin{abstract}
Real estate appraisal is a complex and important task, that can be made more precise and faster with the help of automated valuation tools.  Usually the value of some property is determined by taking into account both structural and geographical characteristics. However, while geographical information is easily found, obtaining significant structural information requires the intervention of a real estate expert, a professional appraiser. 
In this paper we propose a Web data acquisition methodology, and a Machine Learning model, that can be used to automatically evaluate real estate properties. This method uses data from previous appraisal documents, from the advertised prices of similar properties found via Web crawling, and from open data describing the characteristics of a corresponding geographical area.
We describe a case study, applicable to the whole Italian territory, and initially trained on a data set of individual homes located in the city of Turin, and analyze prediction and practical applicability.
\end{abstract}


\section{Introduction}
The correct evaluation of house prices plays a fundamental role in our economy and affects all the participants to the real estate market, including:
\begin{itemize}
\item professional appraisers, who are expert in the evaluation of properties and normally perform \textit{in loco} visits and off-site paper work;
\item real estate appraisal companies, who request, harvest, standardise and verify the work of professional appraisers;
\item financial institutions, needing to (1) set a justified property price prior to offering a mortgage loan or (2) evaluate a property portfolio, e.g. in the context of NPL (Non Performing Loan) management;
\item notaries/solicitors, needing to verify property values prior to guaranteeing the validity of some public transaction (e.g., a deed of purchase or the handling of inheritance issues);
\item home owners and buyers, and real estate agents, wanting to assess the reasonable market price of a property.
\end{itemize}

In relatively recent years, the concept of an Automated Valuation Model (AVM) has emerged in this industry \cite{kok2017big}: an AVM is a software system, often based on online data and resources, that can produce a property evaluation in a semi-automatic way \cite{jaen2002data,Downie:2007,kok2017big}. More recently, Artificial Intelligence and Machine Learning approaches to AVM construction have been adopted \cite{nunez2013artificial,bahia2013data,pow2014applied,ng2015machine,yang2016extended,moosavi2017urban}. This strategy is becoming increasingly useful in two wide areas of business application:
\begin{itemize}
\item Appraisal professionals and companies produce a precise and authoritative evaluation, but with relatively high cost and time requirements - they can reduce such costs and times by using an AVM as a verification system (for the appraisal company) or as a helping tool (for the professional appraiser);
\item Some applications and stakeholders need a quick, or even real time property evaluation, that cannot be produced with the traditional process involving an expert. For example, a street-level bank office may want to immediately propose a draft mortgage offer to an incoming customer, before a formal expert appraisal is available. 
\end{itemize}

When designing an AVM, one must consider the fact that
the appraisal of a real estate property is a very difficult task, due to the high heterogeneity in both structural and geographical data. Moreover the price can be influenced by macroeconomic factors, that obviously change over time. In fact, the creation of a model able to predict real prices needs do deal with several problems caused by the complexity and the dynamics of the real estate market, and to the difficulty of obtaining reliable and objective data.

In this paper, in order to overcome these difficulties, we propose a type of AVM that (1) is adaptive, and uses Machine Learning methods to deal with the complexity and fast-changing characteristics of the real estate market, (2) finds links between diverse sources of data, including open data available on the Web, unstructured data that are obtained via Web Crawling, and private data from previous expert appraisals, and (3) is capable of producing real-time valuations.

A case study is provided, where this proposed Adaptive AVM is applied to the Italian residential real estate market. Three different approaches to feature selection and Machine Learning have been experimented with, yielding surprising results, where many features normally considered important have turned out to be irrelevant. We have used a set of previously appraised residential properties in Turin, Italy, obtaining additional relevant data from heterogeneous Web sources. The results, measured on independent test sets, have shown that the model is predictive and practically effective for the whole Italian territory.

\section{Related work and innovations}

Previous research on Automated Valuation Models for Real Estate has been initially led by so-called "hedonic" models \cite{gt2003,hu2013multivariate, lisi2012estimation, loberto2018potential}. "Hedonic" literally suggests that the buying of the target property, and hence living in it, is a source of "pleasure". The better (and hence more expensive) the property, the higher this specific notion of real estate pleasure, stemming from property characteristics leading to such sensations: a nice view, proximity to services and pleasant life, the presence of an elevator, parking lots, a concierge. We will call all such pleasure-giving characteristics our "hedonic" features. 

In older approaches hedonic features were mainly derived from intrinsic characteristics of the property, e.g. number of rooms and square meters, the number of bathrooms, the floor number. Again, in those traditional approaches, there was generally a linearity assumption - the value $V(P)$ of some property $P$ would depend linearly on the corresponding hedonic features $[f_1(P), ..., f_n(P)]$:

\begin{equation}\label{edonic}
V(P) = \eta + \sum_{1 \leq i \leq n} \beta_i \times f_i(P) + \  \epsilon(P)
\end{equation}

where $\beta_i$ is the "i-th" coefficient, $\epsilon(P)$ is some correction to be applied to this particular property, and $\eta$ is a property-independent correction that applies to some geographical perimeter or application context.

In more recent research, a number of novelties come into place:
\begin{itemize}
\item Non linear and even non-parametric models: we no longer assume the valuation depends linearly on the selected features, as in equation \ref{edonic}. Some studies suggest that this is not in fact the case for real estate valuations \cite{nghiep2001predicting}. In many cases, we just do not know what kind of dependency exists between the features and the sought valuation - we thus follow a non-parametric approach, where the type of regressor is unknown \cite{chopra2007discovering}. In the present study we also follow this approach, and we do not assume linearity nor the correspondence to a particular form of classifier/regressor.
\item Non-hedonic view. Some features are not necessarily "good" or "bad", but rather a part of a more complex analysis. For example, the predominance of foreigners in the neighborhood can lead to higher evaluations for small flats or near a metropolitan city center, while it could have opposed effects in the suburbs and for family houses. We follow this view in this paper, recognizing the complex nature of real estate appraisal.
\item Implementation with AI and Machine Learning. The availability and increased performance of Machine Learning approaches has led to a widespread use of such technologies in AVMs for real estate \cite{NadaiLepri18,jaen2002data,caplin2008machine,ng2015machine,app8112321,gao19}. This includes the use of artificial neural networks \cite{bahia2013data, limsombunchai2004house, nunez2013artificial,tibell2014training,nghiep2001predicting}, decision trees \cite{jaen2002data}, random forests \cite{kok2017big,moosavi2017urban,ceh2018}, gradient boosting \cite{kok2017big} and support vector machines \cite{quanghouse}. In the present paper we have also used Extremely Randomized Trees (Extra Trees) \cite{geurts2006extremely}, that seem to perform well in this context. Finally, multi task learning has been used both for AVMs \cite{gao19} and for DOM (Days on the Market) prediction \cite{zhu2016days}. Other AI techniques may be relevant, such as language classification and even semantic NLP for any text that can be referred to the property neighbourhoods. Image classification and recognition has also been used in the real estate AVM context \cite{ahmed2016house}.
\item Extrinsic features from Web and open data. Not only the features directly related to the target property are important, but also the ones derived from neighbouring amenities and services, as well as linked to totally external information. Space and location-dependent external information has often been found to be important \cite{caplin2008machine,kok2017big}. Neighbouring area information has also been used, including criminality rate, population density and average income, pollution, services and transportation, and the distance from local river banks \cite{NadaiLepri18,bahia2013data,pow2014applied}. In the present study we address this issue in a structured and general way, by defining a notion of "point of interest" (PoI) concerning some subject or service (e.g., transportation, entertainment, sports). Such PoIs are sought for and geographically mapped, based on available open data (see "online resources" at the end of the paper: open data for the Turin municipality - "aperTO", see section \ref{or} below, and nation-wide \cite{agid}, Foursquare, Google Maps, OMI \cite{immobiliare2016manuale}). They are then linked to the target property based on distance and relevance, yielding an organized and comprehensive set of extrinsic features, that are added to the intrinsic features that are present in the appraisal data set\footnote{It should be noted that such PoI indexes can be weighted based on particular valuation contexts or specific target populations, e.g. young people (who could be more interested in public transportation and entertainment), or families (who could be more interested in parks and security)}. 
\item Locally-oriented context. A number of approaches were tailored to a particular geographic area, or evaluated in such local perimeters, e.g. Montreal \cite{pow2014applied}, Beijing \cite{yang2016extended,zhu2016days}, London \cite{ng2015machine}, Los Angeles \cite{caplin2008machine}, Hanoi \cite{quanghouse}, Zurich \cite{moosavi2017urban}, Stockholm \cite{tibell2014training}, Singapore \cite{Ibrahim2005}, Italy (see \cite{lisi2012estimation,fregonara2012value}, as well as the present study), and Slovenia \cite{ceh2018}. This could be seen as a limitation, as if a general methodology was out of hand. Realistically, though, it makes very good sense because (1) valuation practices and regulations differ from one nation to another, (2) different kinds of open data are available and (3) good feature selection is essential and sets of best available features differ geographically. One important contribution of the present paper was to find out that some features obtained from OMI \cite{immobiliare2016manuale} (see section \ref{fs} below) are essential for AVMs targeting the Italian territory.
\end{itemize}

In our approach, we also bring about another novelty, that we believe is not common in the literature. We started from a business and enterprise need, not a consumer-oriented view. For real estate appraisal company, the goal is to obtain a reasonable and expert-supported and validated valuation for some property. We do not target the actual deed of purchase price, nor are we interested in the advertised real estate agency price. As a consequence, we used a data set of expert valuations, not a set of example prices as obtained from Web advertisements, as, e.g., in \cite{loberto2018potential,moosavi2017urban}. We however did use such advertised prices, and obtained them via crawling of specialized Web sites, but only as so-called "comparable" properties (see section \ref{comp} below), that are routinely used by appraisal experts as part of their valuation process.

As a consequence, our approach can be used as a basis for business services to appraisal companies and experts, because it is integrated into their processes and corresponds to their best practices. Moreover, based on our data sets and experiments, we have outperformed the best available results in this context \cite{NadaiLepri18}, representing the current state of the art.

\section{Data set and Open Data acquisition}

For the purpose of this research we have used three different data sources: (1) a corpus of professional and validated property appraisal documents and corresponding data base, (2) geographical and open data obtained from heterogeneous public Web sources, and (3) advertised prices for comparable properties obtained via Web crawling. We discuss such data sources in the next subsections.

\subsection{Data set of available appraisals}

\begin{figure}[!tbp]
\centering
%\setbox0=\vbox{%
%\parindent-4.4em
\includegraphics[width=282pt]{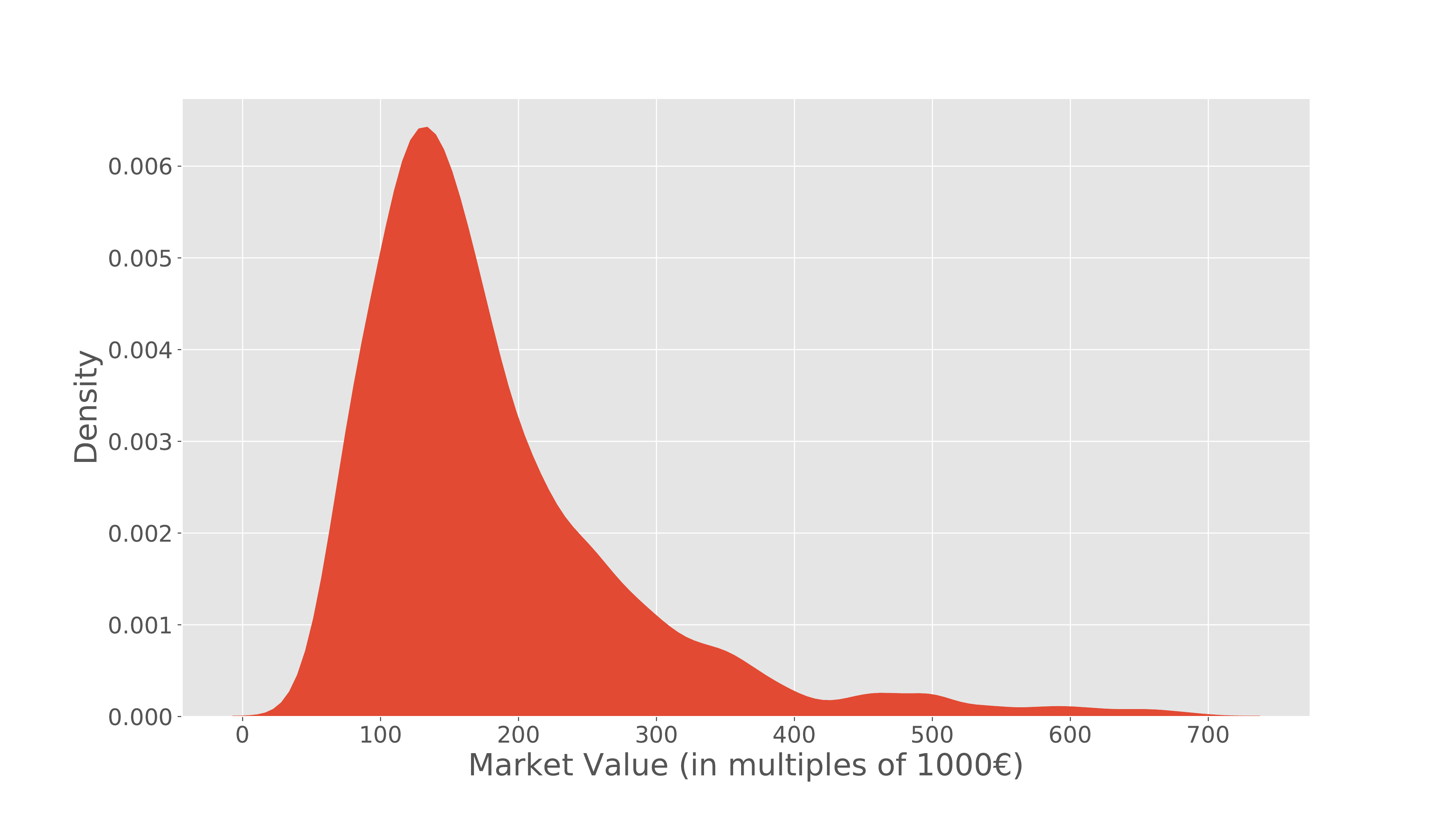}
%\label{price}
%\vspace{-0.25in}
%}\hspace{-2.60in}\makebox[0pt][l]{\box0}
\caption{Valuation distribution in the Appraisal Data Set} \label{fig:valdist}
\end{figure}

The data used to perform the analysis was provided by a multinational real estate appraisal company, with a subsidiary having significant activities in Italy.
The initial data set consisted of 7988 property valuations, performed by professional appraisers, and validated by the company. These properties are all located in the city of Turin, and the valuations were performed between 2011 and 2016. The full set of available information for each property is summarized in Table \ref{dataset}, and the distribution of target valuations in the appraisal data set is shown in Fig. \ref{fig:valdist}.

As a first step, this data set was \textit{anonymized}, removing personal buyer or mortgage application information, as well as bank details.

\begin{table}[ht]
\centering
    \begin{tabular}{ | p{3cm} | p{4,7cm} |}
    \hline
    \textbf{Type of data} & \textbf{Variables} \\ \hline
    Integer & Year of construction, Number of bathrooms, Floor\\
    \hline
    Continuous (square meters) & Surface \\
    \hline
    Boolean (yes/no) & Elevator \\
    \hline
    Ordinal (low, medium, high) & Maintenance status, Quality of Installations, Finishing Quality, View\\
    \hline
    Discrete & Energy Efficiency Classification, Registered Use, Orientation\\
    \hline
    Geographical &  Latitude, Longitude, Address, City Area \\
	\hline
	Target variable (Euros)&  Valuation \\
    \hline
    \end{tabular}
\bigskip
\bigskip
\caption{Information in the available appraisal data set\label{dataset}}
\end{table}

For the most important variables in Table \ref{dataset}, we observe the following:

\begin{itemize}
\item 
Valuation: valuation in Euros, as assessed by a professional appraiser and stored in the appraisal document. This is the target variable, i.e. the value we will want to predict and the output of our AVM when we will use it on new, yet to be evaluated properties. 
\item City Area: Central, Near-central, Larger City Boundary, Suburbs
\item Number of Bathrooms: 0 to 7, mean value = 1.2
\item Surface: 40.4 to 249 square meters, mean value = 94.7
\item Floor: -1 to 11, mean value = 3
\item Registered use: based on the Italian land register ("catasto"), the possible values include, e.g., "residential", "office", "warehouse".
\end{itemize}

This data set contained a wealth of information, that is normally difficult to acquire in such quantity and detail. However, it was too diversified and a \textit{data cleaning and selection} process was performed.  In particular:
\begin{itemize}
\item in order to have an omogeneous data set, we only considered properties with a "residential" registered use, because this was the most common case; by contrast, properties with different registered uses are difficult to compare;
\item  we excluded houses with a surface greater than 250 square meters, as they are rare and belong to a peculiar market sector;
\item we only considered houses with a global valuation between  $20,000$ \texteuro  $\ $ and $700,000$ \texteuro, as the other cases are considered exceptional for the same reasons;
\item we also excluded the valuations of complex properties, as for example houses with a garage, because of their limited number and more complex description structure.
\end{itemize}
After this selection process, the 7988 valuations were reduced to 3983, still a significant number for our practical purposes and prediction targets. 

Finally, we performed a \textit{feature standardisation} activity. For example, some features where normalized so as to have values between -1 and 1, e.g. construction year and floor, so as to make them easier to process in subsequent phases.

\subsection{Geographical and open data}

Using information available on the Web, indexed with the geographical location of the properties, we have extended the features contained in the appraisal data set, with new and useful information. This consists mainly of two data categories: OMI areas and values, and nearby area information.

\subsubsection{OMI areas}\label{omiareas}
\ \\
The Italian Revenue Agency is responsible of the OMI ("Osservatorio del Mercato Immobiliare" - Italian Real-Estate Observatory \cite{immobiliare2016manuale}), see links in Section \ref{or}. For each registered use of properties, the surface of each Italian municipality is divided into different areas, called OMI areas, having homogeneous real-estate characteristics and valuation schemes. 

\begin{figure}[!tbp]
\centering
    \includegraphics[width=252pt]{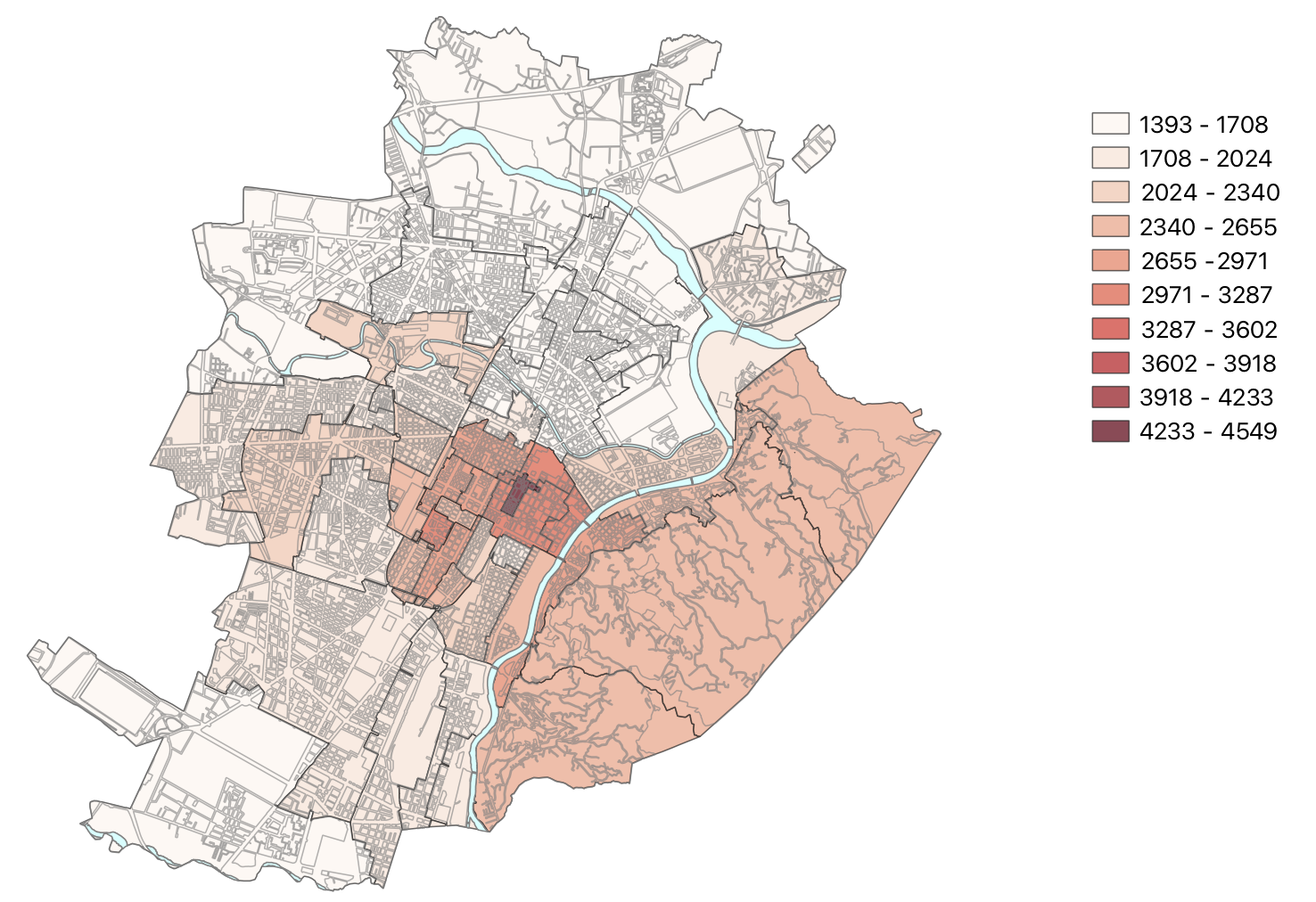}
\caption{Graphical representation of the Turin real estate data set divided by OMI area - average prices} 
\label{price}
\end{figure}

\begin{figure}[!tbp]
\centering
    \includegraphics[width=263pt]{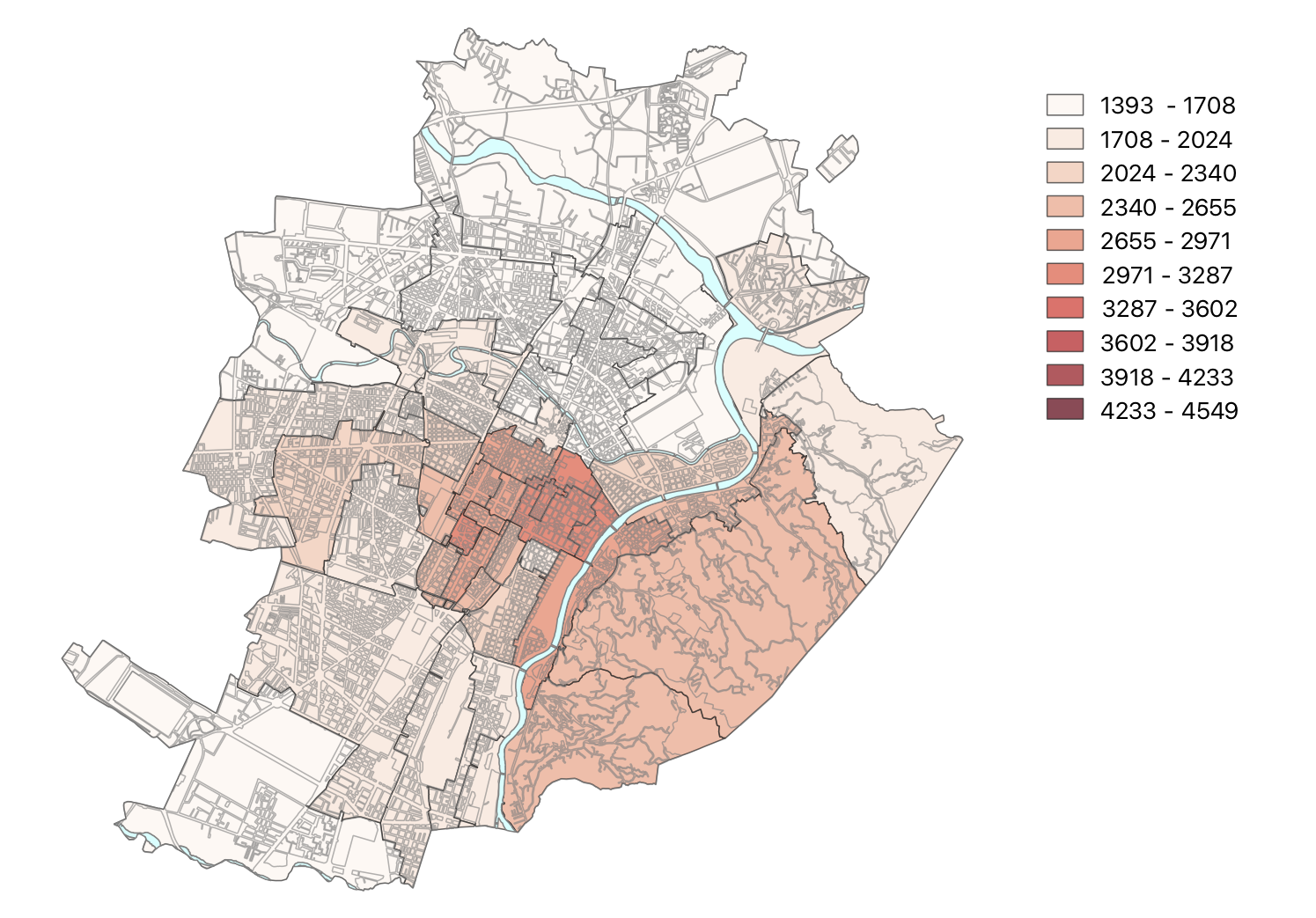}
%}\hspace{-2.78in}\makebox[0pt][l]{\box0}
\caption{Graphical representation of the Turin real estate data set divided by OMI area - predicted prices} \label{pprice}
\end{figure}

Every six months, OMI provides an update of  the price range (minimum and maximum price), for each OMI area. For example, Fig. \ref{price} provides a graphical representation of the 41 OMI areas in the municipality of Turin, with the corresponding average prices. For the purpose of easier comparison, we report next, in Fig. \ref{pprice}, the valuations as predicted by our AVM - as it can be seen the predicted prices are quite close to the actual values. The price ranges in an OMI zone may vary according to the dynamics of the real estate market or to specific changes regarding some geographic location. The OMI area associated to some property may be obtained from OMI, using its geographical coordinates. OMI also provides the formal description of "polygons" that define the OMI areas (see again Section \ref{or} below).

We have followed two different approaches in this research. 

In the first approach (\textbf{OMI names}), we 
used the name of the OMI area as an additional, constructed feature. This has some drawbacks.
First, as previously explained, the price range in a specific OMI area may change over time. This implies that if the model has been trained using valuations performed in a certain period of time, the valuation of a new property in the future could be affected by the changes of price in the corresponding OMI area. Second, the feature will be effective only in the valuation of properties that were located in OMI areas that are represented in the training set. Another issue could be represented by the creation of new OMI areas or the disappearance of old OMI areas over time. Finally, by using the OMI area name, a discrete value, we lose important geographical information, such as the distance between different OMI areas.

In the second approach (\textbf{OMI min/max values}), we introduce as new features the upper and the lower limit of the price range of the OMI area at the time of the appraisal. The idea behind this choice is that of separating the model from specific OMI areas, trying to transform the OMI area name into ordinal features. As the criteria used by OMI for the creation of the price range are the same for each municipality and remain constant over time, this choice allows us to use the model to evaluate properties located in OMI areas not even contained in the training set, and to perform evaluations long after its creation.

%\begin{figure}[!tbp]
%\centering
%\setbox0=\vbox{%
%\parindent-4.4em
%\includegraphics[width=252pt]{Prezzo}
%}\hspace{-2.78in}\makebox[0pt][l]{\box0}
%\caption{Graphical representation of the Turin real estate data set divided by OMI area - average prices} \label{price}
%\end{figure}
%\bigskip
%medskip
%\begin{figure}[!tbp]
%\setbox0=\vbox{%
%\parindent-4.4em

We have performed experiments with both feature construction approaches, and the latter (OMI min/max values) has produced superior results, as discussed later.

\subsubsection{Nearby area information (Points of Interest - PoI)}\label{nearby}

Starting from the geographical position of the property (latitude, longitude), as contained in the appraisal data, we construct new features using open data available from the Web, and related to corresponding surrounding areas (nearby areas). In particular, we built a set of so called "Points of Interest" (PoI). Points of interest correspond to activities and resources that are present in the territory, are geolocalized, and have a potential positive influence on the price of surrounding properties. This corresponds to best practices in real estate appraisal, where experts normally produce valuation reports that include features such as nearby metro and bus stops, schools, and museums. We have grouped our PoIs into 13 categories: Arts, Business\&Services, Entertainment, Food\&Beverage, Healthcare\&Wellness, Instruction, Landmarks, Religious services, Retail, Security, Sport\&Recreation, Transportation, Travel. A data set of PoIs has been created by aggregating information from Foursquare, Google Maps and the Turin Open Data "AperTo" Web Site (see Section \ref{or}).
The idea is to construct a new property feature for each PoI category, by counting the number of PoIs in that category that are within a threshold distance. However, in order to avoid the on/off effect of a strict threshold, we define 4 circles around the property, and associate descending weights to the PoIs falling within these circles, thus giving more importance to the PoIs that are closer to the property. We obtain the following formula, defining the property feature $f_j$ for PoI category $j$:
\begin{equation}
    f_j = \sum_{d_k\leq\frac{r}{8}}1+\sum_{\frac{r}{8}<d_k\leq \frac{r}{4}}\frac{1}{2}+\sum_{\frac{r}{4}<d_k\leq \frac{r}{2}}\frac{1}{4}+\sum_{\frac{r}{2}<d_k \leq r}\frac{1}{8}
\end{equation}
where $d_k$, is the distance of the property from PoI number $k$ in the $j^{th}$ category, and $r$ is a threshold distance, initially set at 1 km. Finally, and additional feature was used, defined simply as the distance from the city center, a special kind of PoI.

\subsection{Comparable properties} \label{comp}
According to best appraisal practices in Italy, the expert's valuation document normally comprises a description and identification of so-called "comparable properties", i.e. properties that are geographically near the target property and have similar characteristics. The price per square meter will then be similar and the professional appraiser will use it as an important starting point in order to reach a final valuation.
Professional appraisers can obtain some such comparable property data from the appraisal company's database of previous valuations. They cannot use purchase deeds from notaries as this is not publicly available in Italy (as it is in France, and, partially, in the UK). Using previous appraisal company valuations will however introduce some bias, as the same experts and the same company standards were used. 

In recent years, appraisers also use real estate offers as advertised on the Web. There are many such publicly available services in Italy, offered to real estate agents as well as private individuals, that allow for sophisticated search interrogations, that may be filtered by, e.g., distance from a specified location, property type, price range, floor, maintenance status. The expert can then download the advertised property description and include a selection of relevant data in her valuation document. The advertised price of such comparable properties will also be used as a reference in setting the valuation of the target property. It must be noted, however, that the advertised price is not a sale price, but rather an upper bound, awaiting further negotiations. The expert will then have to take this simple truth into account when using the advertised price as an input.

In our approach, we have simulated the above expert best practices for comparable data acquisition, while letting Machine Learning do the rest and decide how to use such data. In particular, our system is able to acquire three distinct types of comparable properties:
\begin{itemize}
\item properties that are the target of previous appraisals in the data set: we have such data available in a structured data base, and the corresponding valuation is labeled as a "comparable property valuation price", as it was produced by some appraisal expert some time in the past - the corresponding valuation date is stored and should be taken into account;
\item properties that are cited as comparable in previous appraisal documents: this is available in the expert's text document, and not in the data base, so we were not able to extract it easily - we did not use this at the present time;
\item properties that are advertised for sale on the Web.
\end{itemize}

For the latter category of comparable properties, we simulated the behaviour of the expert using a controlled Web Crawling strategy, where a limited number of properties is sought near the target property. If too many properties are found, the distance threshold is reduced and additional filters are set, with the purpose of finding properties that have similar characteristics (e.g. floor and maintenance status). If the demonstrator is scaled up and used in a production and commercial service, a number of issues will have to be addressed, including legal concerns about the use of a robot to retrieve possibly proprietary information (though actually publicly available on the Web). Some technical issues should also be analyzed, such as dealing with anti-automation (e.g., Captchas), and robot detection/classification \cite{Perna:2018}. Finally, frequent Web format changes in real estate portals will require manual Crawler adaptation - again, legal issues should be addressed here.

The demonstrator is now able to retrieve a set number of comparable properties as described above, with corresponding attributes and advertised or valuated price. The result of the process is a new constructed feature for the target property: the average price per square meter of comparable properties.

\section{Machine Learning methodology}
We will now describe how the described data set was used, in order to train our AVM. First we will describe three different approaches that we have used for feature selection, and subsequently we will describe how the data set was partitioned and which Machine Learning algorithms were used.

\subsection{Feature selection}\label{fs}
The data set described in the previous section is complex and involves significant amounts of correlated information. We have followed three distinct approaches to the use of such data, by selecting different subsets of the available features:
\begin{enumerate}
\item \textbf{Hedonic features}. In this approach, we select features that seem naturally influence the value of the property, in either positive or negative ways (so called "hedonic" features \cite{limsombunchai2004house,lisi2012estimation}). For example the presence of an elevator or a good maintenance status will be associated to higher valuations, while distance from the center negatively influences the appraisal. Specifically, we have used: (1) all the features contained in the appraisal data set (Table \ref{dataset}), (2) the POI features for the categories described in Section \ref{nearby} (e.g., transportation, entertainment, etc.), (3) the distance from the city center, and (4) the OMI area name, as defined in Section \ref{omiareas}.
\item \textbf{OMI-centered features}. By using hedonic features in the learning step, with different algorithms and hyper parameters, it turned out that many features were irrelevant and seldom used by the output regressors. We then adopted a new approach, with a substantially reduced feature set, including: (1) OMI minimum and maximum price per square meter valuations for the OMI area where the property is located (OMI-min, OMI-max), (2) surface in square meters. The identification of the correct OMI area is obtained from "polygons" that are made available by OMI (see Section \ref{or}), and based on the property location (latitude and longitude).
\item \textbf{OMI-centered features and comparable prices}. In this case, we used the same features as in the previous approach, with the addition of the average price per square meter of comparable properties.
\end{enumerate}
As it will be evident from the results described in the next section, the latter approach provides the best prediction.

\subsection{Learning step}
The available data set was initially shuffled, so as to avoid order dependence and position-based regularities. Then it was partitioned into three separate subsets: \\ 
\\
learning set (70\% $\implies$ 2789 valuations)\\
validation set (15\% $\implies$ 596 valuations)\\
test set (15\% $\implies$ 596 valuations)\\
\\
where the learning set is used to train a regressor, the validation set is used to choose the best hyper-parameter values needed by the learning algorithm, and an independent test set is used to evaluate and report the results.

The learning task is a supervised regression problem, where the target feature to be predicted is the property valuation. We have experimented, depending on the feature selection method, with up to six Machine Learning regression algorithms: K nearest neighbour (KNN), Random Forests, Extremely Randomized Trees (Extra Trees) \cite{geurts2006extremely}, Gradient Boosting, Adaboost, and Bagging. The hyper-parameters tuned with the validation set can deeply influence the performance of the above regressors,  for example the value of K in KNN,  or the number of trees and the maximum tree depth in a random forest.

\begin{figure}[!tbp]
\centering
\includegraphics[width=240pt]{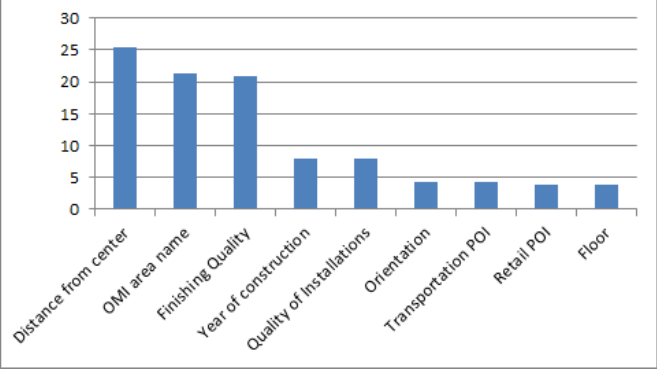}
\vspace{-0.05in}
\caption{Relative importance of the Hedonic feature set, using the Random Forest regressor (best 9)} \label{fimpRF}
%\end{figure}
\bigskip
\medskip
%\begin{figure}[!tbp]
\centering
\includegraphics[width=240pt]{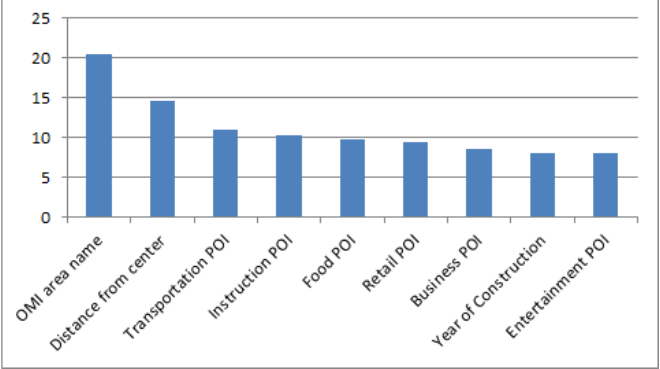}
\vspace{-0.05in}
\caption{Relative importance of the Hedonic feature set, using the Gradient Boosting regressor (best 9)} \label{fimpGB}
\end{figure}

Some of the above learning algorithms were also used for an analysis of the feature set. For example, with Random Forests, Extra Trees and Gradient Boosting, it is possible to measure the relative importance of features, based on how often they occur in the learned regressors. For the Hedonic feature set (as defined in section \ref{fs}), we obtained the rankings described in Figures \ref{fimpRF} and \ref{fimpGB}, for the Random Forest and the Gradient Boosting regressors, respectively\footnote{An even more important feature consists in the property surface, in square meters. This is not shown in Figures \ref{fimpRF} and \ref{fimpGB}, because it is out of scale, and also because it can be excluded by using the price per square meter as a target regression feature}. In the figures, the ranking values reported on the y axis sum up to 100 for the 9 most important features, that are reported on the x axis.

When using other algorithms, we obtained partly similar rankings - in particular, the OMI area name always ranks first or second. This prompted us to a radical change in addressing feature importance. In fact, features that we initially thought to be relevant, such as the POI indexes for Arts and Security, were rarely used. This is maybe due to the fact that the target price is an expert valuation, and not the real deed of purchase value. However, we had to stick to this prediction problem, as it was the only possible, and face the fact that the Hedonic feature set was far from perfect. The presence of many irrelevant features could have let to overfitting phenomena. At the same time, all learning algorithms gave great importance to OMI areas and distance from the city center, that are obviously correlated.

Further analysis highlighted the fact that OMI areas, in the Hedonic feature set, were coded by names. As a consequence they do not carry any geographical properties and correlations, e.g. proximity between different OMI areas. To correct this issue, we found that it was possible to obtain from OMI, as a payed service, the mininum and maximum property price in any given area, as evaluated by OMI itself (OMI-min and OMI-max). By training regressors after adding these two features to the hedonic feature set, it turned out that they are by far the most important features, and significantly superior even to the OMI area name. 

As a consequence, the use of the regressors as feature ranking tools, based on the training set, led us to the addition of the OMI-min and OMI-max features, and to the use of the reduced feature sets defined in Section \ref{fs} (OMI-centered, and OMI-centered with comparables). These feature sets are associated to the best results and show the best accuracy on the test set, as discussed next for each of the above-cited learning algorithms.

\section{Results}
We now describe the results, as measured on the test set, for the three different feature sets. The results for the Hedonic feature set are reported in Table \ref{res1}, whereas the results for the reduced, OMI-centered feature sets are reported in Tables \ref{res2} and \ref{res3} and in Figures \ref{omiscatter} and \ref{omicscatter}. We also report results on a totally separate out of sample test set in Fig. \ref{outs}, with different geographic areas and time intervals. 

\subsection{Results for the hedonic feature set}

For the hedonic feature set, with the addition of the OMI area name, we mainly consider the mean error (ME):
\begin{equation}
ME =
\sqrt{1/n\sum_{[x_i,y_i]\in TS} (y_i-\hat{y_i})^2}
\end{equation}
where $\hat{y_i}$ is the predicted value, and $[x_i,y_i]$ is an example in the test set $TS$, with the target feature being the property valuation $y_i$ and $|TS|=n$. This is expressed in thousands of Euros, and the results are reported in Table \ref{res1}, where the Mean Square Error ($MSE = ME^2$) and the coefficient of determination $R^2$ are also shown.

\begin{table}[ht]
\centering
 \begin{tabular}{ | p{3cm} | p{1cm} | p{1cm} | p{1cm} | p{1cm} |}
    \hline
    \textbf{Algorithms} & \textbf{ME} & \textbf{MSE} & \textbf{$R^2$} \\ 
	\hline
	Bagging Regressor & 46.4 & 2155 & 0.774\\
    \hline
    AdaBoost & 52.8 & 2787 & 0.707\\
    \hline
    Gradient Boosting & 43.9 & 1935 & 0.797\\
    \hline
    K-Nearest Neighbor & 54.4 & 2964 & 0.689\\
    \hline
    Extra Trees & 41.6 & 1738 & 0.810\\
    \hline
    Random Forest & 43.1 & 1864 & 0.807\\
    \hline
    \end{tabular}
\bigskip
\caption{Results for the hedonic feature set\label{res1}}
\end{table}

Since the mean property value in the data set was 189.400 Euros, the mean errors in Table \ref{res1} cannot be considered a very good result, though it can be of aid to an professional appraiser as a rough indication and a starting point. Extra trees and Random Forests seem to perform best with this feature set.

\subsection{Results for the reduced feature set}

For the reduced feature set (OMI-centered), and for the different learning algorithms that were previously listed, we have obtained the results shown in Table \ref{res2}.

\begin{figure}[!tbp]
\centering
\includegraphics[width=240pt]{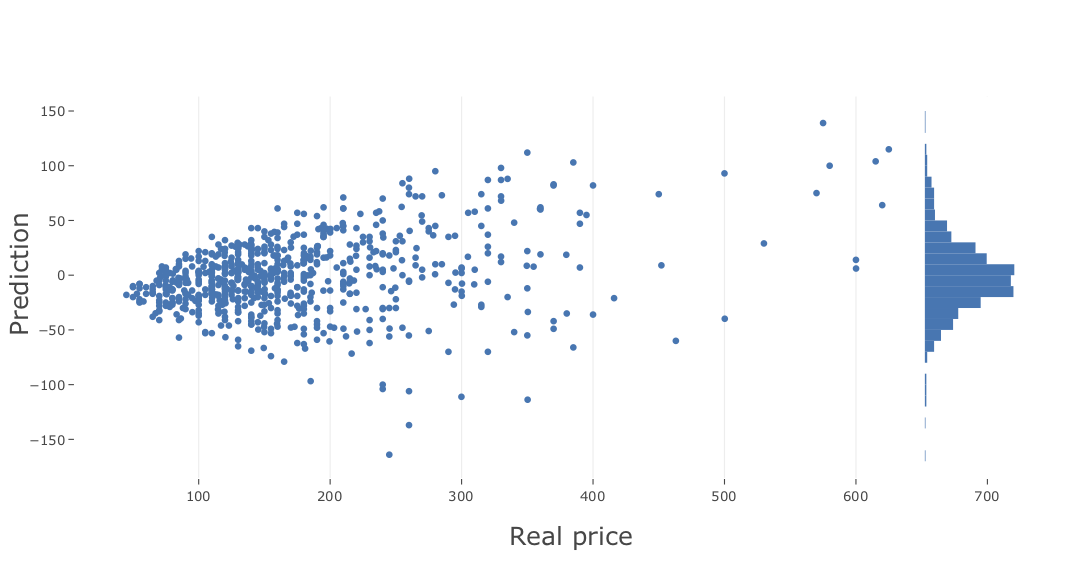}
%\vspace{-0.05in}
\caption{Results for OMI-centered features (random forest)} \label{omiscatter}
%\end{figure}
\bigskip
\medskip
%\begin{figure}[!tbp]
\centering
\includegraphics[width=240pt]{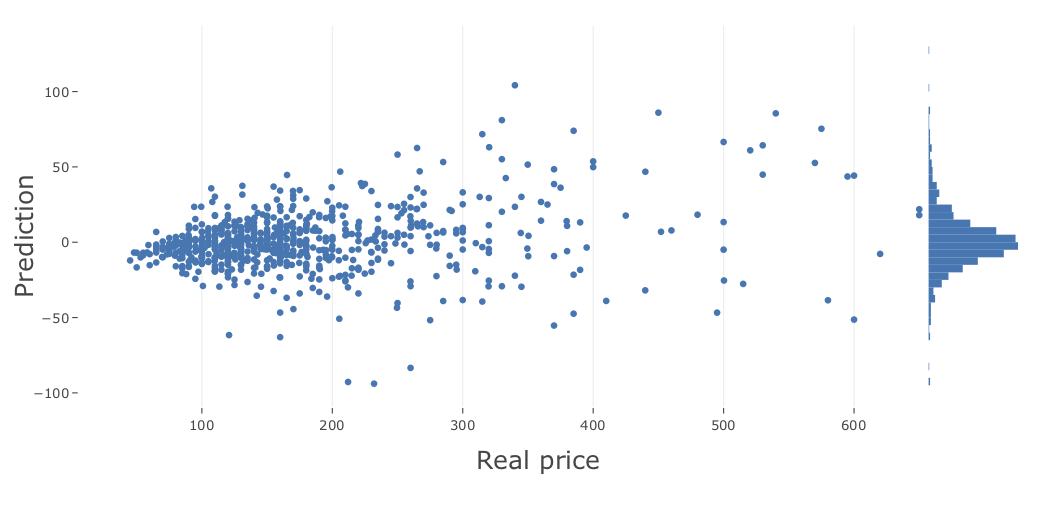}
%\vspace{-0.05in}
\caption{Results for OMI-centered features plus comparables (random forest)} \label{omicscatter}
\end{figure}

\begin{table}[ht]
\centering
 \begin{tabular}{ | p{3cm} | p{1cm} | p{1cm} | p{1cm} | p{1cm} |}
    \hline
    \textbf{Algorithms} & \textbf{ME} & \textbf{MSE} & \textbf{$R^2$} \\ 
	\hline
	Bagging Regressor & 37.74 & 1425 & 0.838\\
    \hline
    AdaBoost & 39.29 & 1543 & 0.824\\
    \hline
    Gradient Boosting & 47.41 & 2248 & 0.744\\
    \hline
    K-Nearest Neighbor & 46.24 & 2139 & 0.757\\
    \hline
    Extra Trees & 34.84 & 1214 & 0.862\\
    \hline
    Random Forest & 35.79 & 1281 & 0.854\\
    \hline
    \end{tabular}
\bigskip
\caption{Results for the OMI-centered feature set\label{res2}}
\end{table}

This is significantly better than the hedonic feature approach. The best results are obtained for Bagging, Random Forests and Extra Trees. For these algorithms, after introducing comparable values, the results improve further, as shown in table \ref{res3}. The mean error is now around 21 k\texteuro, that has been considered acceptable by the domain experts, leading to a tool that would be very useful in the appraisal process. 

A comparison of results for OMI-centered features, with and without comparables, can be seen in different detail in Figures \ref{omiscatter} and \ref{omicscatter}. Each dot in these scatter plots represents a real estate property in the test set, with the projection on the x axis representing its expert-valuated price, and the projection on the y axis representing the error in our AVM prediction. It is immediately evident that the accuracy of the AVM with comparables (Fig. \ref{omicscatter}) is significantly better, making the corresponding Web Crawling activity worthwhile. This can also be observed on the concentration graph on the right of both figures.

\begin{table}[ht]
\centering
 \begin{tabular}{ | p{3cm} | p{1cm} | p{1cm} | p{1cm} | p{1cm} |}
    \hline
    \textbf{Algorithms} & \textbf{ME} & \textbf{MSE} & \textbf{$R^2$} \\ 
	\hline
	Bagging Regressor & 21.61 & 467 & 0.957\\
    \hline
    Extra Trees & 21.12 & 446 & 0.959\\
    \hline
    Random Forest & 21.48 & 461 & 0.958\\
    \hline
    \end{tabular}
\bigskip
\caption{Results for the OMI-centered feature set with the addition of comparables\label{res3}}
\end{table}

\subsection{Test out-of-sample}

We have also used another small data set which contains the valuations of 58 properties located in different Italian cities, that were performed from year 2012 to year 2016. By contrast, the Turin-area valuations in our Data Set, that were used to train the model, were performed between 2015 and 2016. 

\begin{figure}[!tbp]
%\setbox0=\vbox{%
%\parindent-4.4em
\centering
\includegraphics[width=272pt]{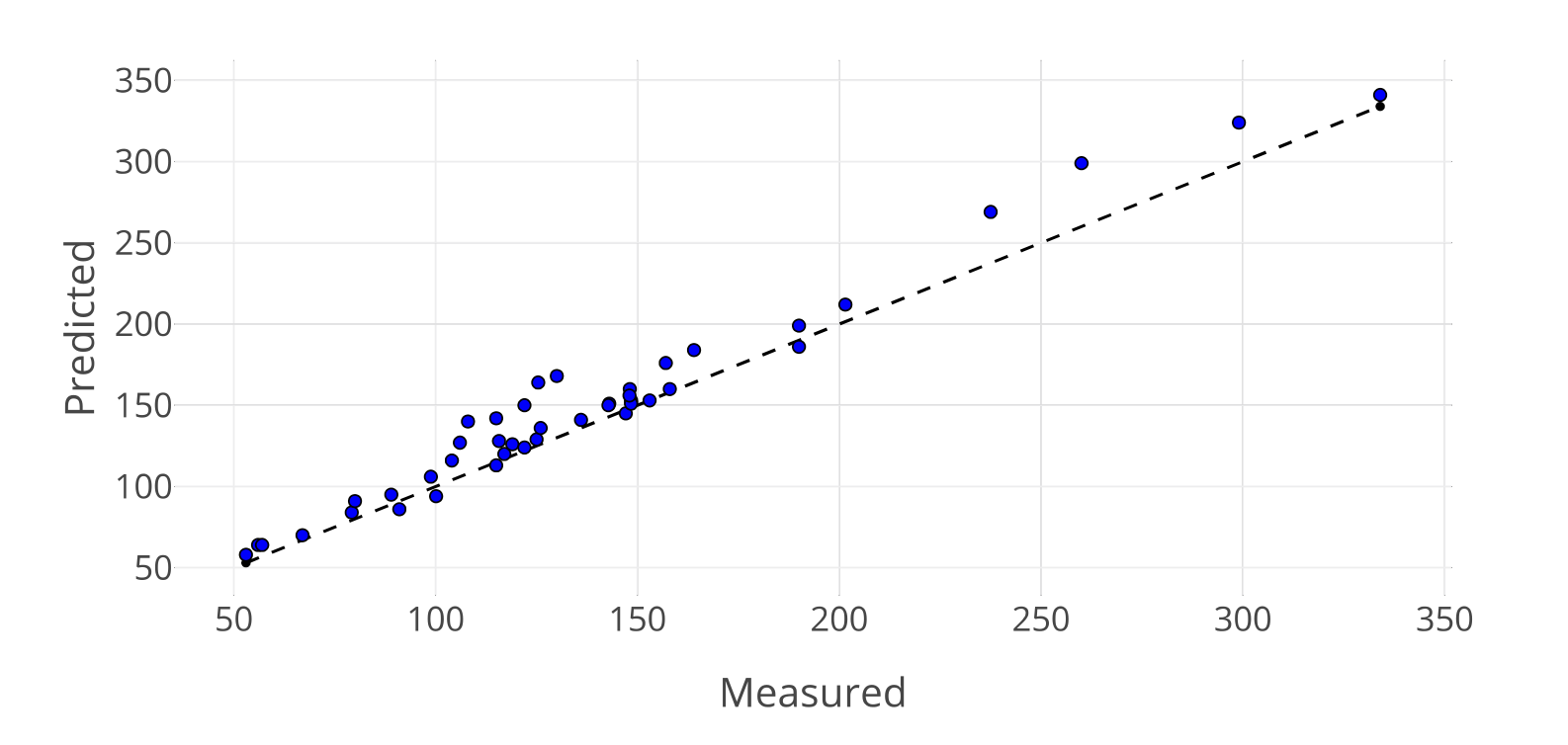}
%}\hspace{-2.70in}\makebox[0pt][l]{\box0}
\vspace{-0.1in}
\caption{Out of sample test (prices in multiples of 1,000 \texteuro)} \label{outs}
\end{figure}

For this totally new and uncorrelated data set, we obtained a Mean Error of 17,000 Euros, even lower than what was observed on the test set with OMI-centered features and comparables (Table \ref{res3}). This confirms that the model learned using the Turin data set is predictive, and can be used on new property evaluation tasks that are different in terms of both time and space. The results are shown in Fig. \ref{outs}, where the dots' projection on the y axis represent the AVM prediction for the corresponding property valuation, while the projection on the x axis is the expert's evaluation. The results are very good, and closely approximate the "perfect" prediction, represented by the dotted straight line.

The implemented demonstrator is now able to run automated valuations for a property located anywhere in Italy, as a general function was implemented to compute the OMI area name based on the property's longitude and latitude and on the area polygon obtained from OMI (available as open data). Web crawling for comparables was also partially automated for the whole Italian territory.

\section{Conclusions}

This paper presents a complex data acquisition and learning system, that brings about three main contributions:
\begin{itemize}
\item we have developed a methodology for acquiring relevant real estate property features from the Web and open data, correlating them to other existing intrinsic features that are normally available in expert valuation documents and appraisal data bases;
\item we have shown that, using such features, it is possible to predict the value of some property with a limited error rate;
\item we have developed an AVM tool that uses this data acquisition of Machine Learning methodology, and may be a valid help for professional appraisers and for appraisal companies that need to validate expert documents or evaluate real estate portfolios.
\end{itemize}
The approach followed in this research should be refined in order to obtain an even more performing feature set, and should be validated on larger and international data sets. Further exploitation should then address open issues, such as technical and legal aspects of Web Crawling in search of comparable property prices, and partial automation of the appraisal document preparation process.

\section*{Competing interests}
  The authors declare that they have no competing interests.

\section*{Author's contributions}
    Conceptualization: GR, DP, FB; Data curation: RB; Formal analysis and methodology: DP, GR, FB; Writing: FB, RB, GR, DP.

\section*{Acknowledgments}

We would like to thank Netatlas s.r.l and Certimeter Group for providing the experimental set up and the possibility of using the data set for the purposes of this research. We would also like to thank Netatlas s.r.l. and Fondazione CRT (Lagrange project) for funding part of this project.

\section*{Online Resources}\label{or}

\textbf{OMI (Italian Real Estate Observatory):}\\
\url{https://www.agenziaentrate.gov.it/wps/content/nsilib/nsi/schede/fabbricatiterreni/omi}\\
\textbf{OMI area perimeters:} \\
\url{https://wwwt.agenziaentrate.gov.it/geopoi\_omi/index.php}\\
\textbf{Open Data Torino:} \url{http://aperto.comune.torino.it/}\\
%\textbf{google maps:}
\textbf{Foursquare:} \url{https://www.foursquare.com/}\\
\textbf{Google Maps APIs:}\\ \url{https://developers.google.com/places/web-service/intro}

\bibliographystyle{unsrt}

\vspace{1ex}

\listoffigures
\listoftables

\end{document}